 \documentclass[preprint,3p,times,twocolumn]{elsarticle}
\usepackage{amssymb}
\usepackage[hyphens]{url}

\journal{The journal of network and computer applications}

\begin{document}

\begin{frontmatter}


%% \title{Detecting fraudulent activity in a cloud using privacy-friendly data aggregates\tnoteref{label1}}

\title{Detecting fraudulent activity in a cloud using privacy-friendly data aggregates}


\author[label1]{Marc Solanas}

\address[label1]{marc@solanas.cat}

\author[label2]{Julio Hernandez-Castro}

\address[label2]{J.C.Hernandez-Castro@kent.ac.uk}

\author[labeld]{Debojyoti Dutta}

\address[labeld]{dedutta@cisco.com}

\begin{abstract}
More users and companies make use of cloud services every day. They all expect a perfect performance and any issue to remain transparent to them. This last statement is very challenging to perform. A user's activities in our cloud can affect the overall performance of our servers, having an impact on other resources.\\
We can consider these kind of activities as fraudulent. They can be either illegal activities, such as launching a DDoS attack or just activities which are undesired by the cloud provider, such as Bitcoin mining, which uses substantial power, reduces the life of the hardware and can possibly slow down other user's activities.\\
This article discusses a method to detect such activities by using non-intrusive, privacy-friendly data: billing data. We use OpenStack as an example with data provided by Telemetry, the component in charge of measuring resource usage for billing purposes. Results will be shown proving the efficiency of this method and ways to improve it will be provided as well as its advantages and disadvantages.

\end{abstract}

\begin{keyword}
OpenStack \sep Ceilometer \sep Cloud Computing \sep Machine Learning \sep Classification \sep Security \sep Networks \sep Intrusion \sep Random Forest \sep Bitcoin Mining



\end{keyword}

\end{frontmatter}


\section{Introduction}

Cloud computing used to be tech leaders' dream: a paradigm where computing becomes a service. It allows companies and users to use as much compute power – or storage – as needed, paying only for what is used. In this way, using one hundred computers during one hour costs the same as using one single computer for one hundred hours \cite{armbrust2010view}.\\
Users of cloud computing have many different needs. Some use a website, for example a social network, some perform mathematical calculations and some use a virtual machine to deploy a production web application, among many other uses. They all require different levels of abstractions. A user of a webmail client does not need to know what operating system the server is using while someone who wants to create a website might not know how to manage a webserver.\\
The most common paradigms these days are SaaS, Software as a Service; PaaS, Platform as a Service; and IaaS, Infrastructure as a Service \cite{peng2009comparison}. If we look at them from the point of view of security, in the first two cases, the user has limited capabilities to conduct fraudulent activities due to the lack of power he has. There are always security risks --- no one can deny this ---, but the paradigm that is most useful for fraudulent activity is the last one: IaaS, which normally offers complete administrator permissions to the user making it essential to have an intrusion detection system in place, to avoid other users being affected by the behavior of the fraudulent user \cite{secissues}.\\
Usually, users in a cloud are completely isolated, meaning that it is not common for one user to get access to data owned by someone else or directly interfere with other workloads. However, most clouds are configured to be flexible, allowing a user to use more than the assigned resources at some points, to account for pikes in use, or just because they are not being used. In addition, Ethernet links are shared amongst all the workloads, which make them an easy target. The more bandwidth someone uses, the slower the other workloads will run.\\
In addition, there are activities, which might appear as non-fraudulent to the user but are undesired for cloud providers. I am talking about Bitcoin mining \cite{nakamoto2008bitcoin}, or Litecoin mining \cite{litecoin}, which is becoming increasingly popular and harder to do. Mining cryptocurrencies uses much higher power consumption and, furthermore, it reduces the lifespan of our software \cite{kamplbitcoin}. In addition, there is the legal aspect of creating digital money using someone else's infrastructure. As a consequence, cryptocurrency mining is something most cloud providers would like to avoid.\\
Most intrusion detection systems (IDS) nowadays require network packet inspections \cite{snort}, invading a user's privacy. In the age of Internet surveillance, it is understandable that users do not feel comfortable having all their data inspected by a third party application. Is there a way to detect fraudulent activity while respecting a user's privacy? A way of doing this would be to use data aggregates, which do not give a lot of detail, such as CPU usage or the number of outgoing packets in a closed interval, to perform a first classification. Data aggregates as a privacy-friendly method to collect data has previously been used by many others such as \cite{he2007pda}. In case a fraudulent activity is suspected, then a more in-depth method can be used. This way allows users who run regular workloads to keep their privacy while detecting suspicious activity.

\subsection{Objectives}
This article details a method to collect samples of data from an OpenStack cluster \cite{openstack}, featuring regular workloads and fraudulent ones. With this data we are to try different classification algorithms such as SVM or random forests in order to find the one offering the best results.\\
We will try to classify 5 types of jobs: a regular workload, hadoop, which can use resources in many different ways; a very CPU-intensive job, simulating mathematical calculations; an internal DDoS attack, from one virtual machine to another inside our cloud; a physical network failure, to see if I can distinguish it from a regular job; and cryptocurrency mining, I'll get data of both Bitcoin and Litecoin mining.\\
Once initial results are shown and an algorithm chosen, we will proceed to discuss ways to improve the results by making use of a meta-classifier, creating a proof-of-concept that, using the existing data, simulates how it would work in a real cloud, in real time.\\
In conclusion, we define a method to detect fraudulent or suspicious activity in a cloud environment using metrics, which are normally collected for billing purposes, which do not offer too much insight on the user's activities but it can detect an activity as suspicious and trigger an alarm for further investigation.

\section{OpenStack}
OpenStack, born as a joint project of Nasa and Rackspace in 2010, is an open-source Cloud Computing platform for public and private clouds. Its main goal is to enable companies of any size to have their own private cloud at a low cost, feature rich and with a vibrant community developing it all over the world \cite{openstack}. Before the release of open source cloud operating systems, companies had two options: they either bought clusters or rely on external companies who provided a cloud, such as Amazon. With OpenStack, or other alternatives, such as CloudStack \cite{cloudstack}, companies can use their existing infrastructure for an in-premise, private cloud, which grants them a much higher security.\\
OpenStack offers a set of separate components, which make it very easy to scale, offering different services such as compute, bock or object storage, network, databases or orchestration as a service (look at the figure below). The following sections take a look to the most important of these components.

\begin{figure}[htb]
\begin{center}
\includegraphics[scale=0.6]{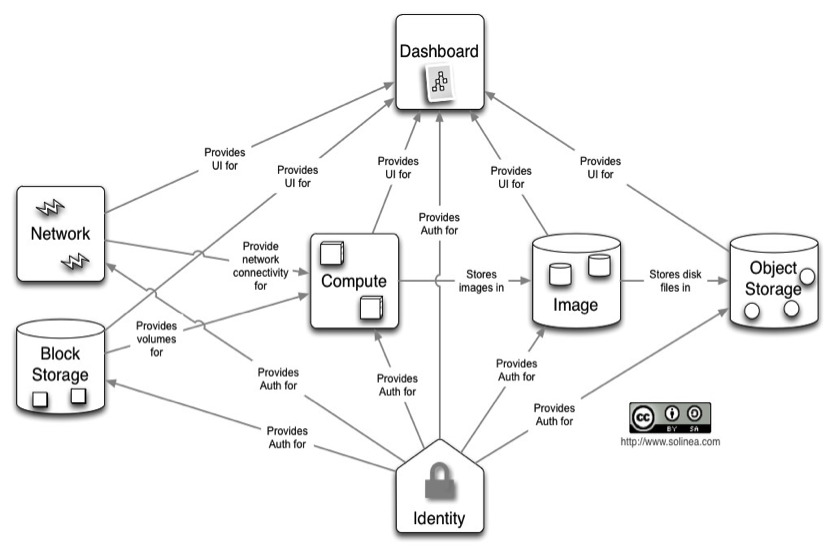}
\caption{OpenStack architecture. Source: OpenStack Foundation.}
\end{center}
\end{figure}

\subsection{Main Components}
\subsubsection{Compute (Nova)}
Nova is the subproject that takes care of providing compute power as a service. It can be plugged to one or several hypervisors such as Qemu \cite{qemu} or KVM \cite{kvm}. It consists of a controller node, which uses an API to take requests, and several compute nodes, which are the ones where the virtual machines will reside and whose resources they will use \cite{openstack-projects}.\\

Compute nodes can be added at any point very easily, as well as removed, making Nova highly available, with a single failure point, which is the controller, and very easy to scale. The controller node is the only one who has knowledge of all the computes and, based on the resources they have, their availability and another increasing number of constraints, it decides where to deploy each new virtual machine.

\subsubsection{Glance (Image)}
Glance provides the user with a RESTful API to store, discover and serve virtual machine images. It can also store volumes, which can be used as a template for new ones \cite{openstack-projects}.\\
These images can be stored in a wide variety of file systems, either in block storage or object storage file systems such as the one managed OpenStack Swift.

\subsubsection{Cinder (Block Storage)}
Cinder takes care of providing block storage as a service. In reality, it started as a fork of Nova, which was modified to manage virtual volumes instead of virtual machines. These volumes can, again, feature any file system suitable to the user, such as Ceph \cite{openstack-projects}.

\subsubsection{Keystone (Identity)}
The Keystone subproject manages authentication, tokens and policies for the entire OpenStack ecosystem by providing a RESTful API used by projects to control access permissions as well as to discover and provide access to other services in OpenStack, which are registered in Keystone \cite{openstack-projects}.

\subsubsection{Neutron (Network)}
Neutron, previously known as Quantum, provides network as a service to any resource created and managed by other components. For example, it can create different vlans to isolate several virtual machines from each other, as well as creating virtual network components such as routers. Furthermore, with recent updates, it also provides Firewall as a Service and Load Balancer as a Service \cite{openstack-projects}.

\subsubsection{Swift (Object Storage)}
Swift is the project in charge of highly available, distributed object storage. Recently, a lot of companies are using Swift together with Ceph, an open source distributed file system by Inktank, in order to provide higher availability \cite{openstack-projects}.

\subsubsection{Ceilometer (Telemetry)}
The most relevant component for this article is Ceilometer, now known as OpenStack Telemetry. It was first meant to be a metering component for OpenStack. However, many other applications have been suggested, some of them are very interesting. The reason why so many different applications have arisen is because Ceilometer's job is to report usage metrics, data from each resource in OpenStack at regular intervals \cite{openstack-projects}. The following section will describe Ceilometer with some more detail.

\section{Infrastructure and operating system}

The OpenStack cluster used to collect the data for this article was a set of 3 Intel NUCs D54250WYKH1 \cite{nuc} provided by Cisco Systems. The NUCs have 2 network cards each, an Ethernet one, connecting them through a switch, and a wireless one, giving them Internet access.
In terms of storage, they have 2 solid state drives each, a 250GB one for the operating system and OpenStack installation and a 1TB one to be used by Cinder for object storage. In addition, each has 16GB of ram.\\
The operating system used is Ubuntu 14.04 amd 64 + OpenStack IceHouse. The Ethernet nic has been used as the flat interface and the wireless one as the public interface. Neutron, the networking component of OpenStack has not been used in favor of nova's legacy networking, which was simpler to install and, because of the types of data collected, did not alter the result.\\
One of the NUCs was used as both a controller and a compute node, and the other two acted as compute nodes.

\section{Data collection}
\subsection{Justification of the data collected}
As previously mentioned, we plan to use data aggregates of several common metrics, instead of performing packet inspections or looking for flow patterns like many intrusion detection systems in the market currently do, such as Snort \cite{snort}.\\
The reason for this is to offer a higher privacy to the user. In a society where concern for privacy is increasing exponentially, we need more solutions that offer security without sacrificing a user's privacy. A way to respect it is to make the data anonymous by hiding detailed patterns. We collect slices of usage, which show us rates of change of different metrics over the last few seconds. Similar rates of change probably mean similar activity.\\
In our case we use five seconds aggregates of 3 of the most common metrics available in all systems: CPU, disk and network.

\subsection{Types of data}
One of the initial goals was to use a small number of metrics. Before using each metric I have transformed it to show the rate of change of the metric during the current interval \cite{ceilometer-transformer}. The metrics are:
\begin{itemize}
\item CPU\_util: in tan percent.
\item Disk.read.requests.rate: in requests/second.
\item Disk.read.bytes.rate: in bytes/second.
\item Disk.write.requests.rate: in requests/second
\item Disk.write.bytes.rate: in bytes/second.
\item Network.incoming.bytes.rate: in bytes/second.
\item Network.incoming.packets.rate: in packets/second.
\item Network.outgoing.bytes.rate: in bytes/second.
\item Network.outgoing.packets.rate: in packets/second
\end{itemize}
All this data is collected per each virtual resource in our cluster – virtual machines, volumes, virtual routers, etc. –, every 5 seconds. In this way, we can easily remove data from resources we do not need such as other users' virtual machines.
It is worth pointing out the reason why both network incoming/outgoing bytes and packets are collected. The reason for this is to enable the classifier to distinguish between applications that send few, big packets, instead of many, small ones, allowing it to create a ratio with the number of packets and size.\\
This pipeline is configured in the Ceilometer pipeline.yml file located under /etc/ceilometer in each compute node. Once it has been modified, you need to restart the ceilometer-agent-compute process in all the servers. The file should instantly appear in the specified path and data samples will be appended to it every 5 seconds.\\
The next step is to run an experiment, such as the ones described below, to generate data.

\subsection{Experiments}
In order to proof that this method can not only detect fraudulent activities but also classify them, to a certain extent, I decided to run several experiments including several types of fraudulent activities, infrastructure failures and regular workloads.

\subsection{General workloads}
Most the common workloads that run in our clouds have very different fingerprints \cite{top-uses}. Some will require little activity, while some others will use a lot of resources.\\
In order to run experiments that represent as many common workloads as possible, I decided to use Hadoop and run a benchmark suite by Intel, called HiBench \cite{hibench}, which runs several hadoop jobs in a sequence, containing benchmarks stressing certain resources, such as CPU or disk, but also regular algorithms such as PageRank.\\
In addition, in order to make sure I can distinguish between cryptocurrencies and very CPU-intensive workloads, I also ran extra experiments using the Linux Stress tool \cite{stress}.

\subsection{Fraudulent activities}
\subsubsection{Internal DDoS attack}
The first fraudulent activity I created data of is an internal DDoS attack. By internal, I mean that it originates from virtual machines within our cloud and it is aimed to another virtual machine within our cloud. OpenStack offers resource over allocation by default, when possible, so a hacker could perform such an attack in order to attempt to slow down resources belonging to other users, impacting in their business operations and resulting in a potential decrease in income.\\
The attack was a Ping flood attack \cite{ddoss}. I set it up by creating a custom Ubuntu image that, upon boot, would monitor a remote file – in a different cluster – waiting for a victim's ip. Once an ip was set in this remote file, it would create one hundred threads that would ping the victim, with the maximum allowed ICMP packet size, 65,535 bytes.\\
The experiment consisted of 50 virtual machines that attacked another machine in the same cluster using its IP. While running it, I could clearly notice a decreased overall speed in the cluster and, in some cases, I received timeout errors when trying to access virtual webservers hosted in one of the nodes affected by the attack.

\subsubsection{Cryptocurrency mining}
Cryptocurrencies have become increasingly popular in recent years, starting with the release of Bitcoins reaching a price of \$1242 per coin in November 29th 2013 (see figure below). With this growth comes a growth in the difficulty of mining these currencies \cite{kroll2013economics}. It is not unexpected, as a consequence, that users try to use cloud environments to mine them, using a lot of resources, and decreasing the life of the hardware in the data center. 

\begin{figure}[htb]
\begin{center}
\includegraphics[scale=0.6]{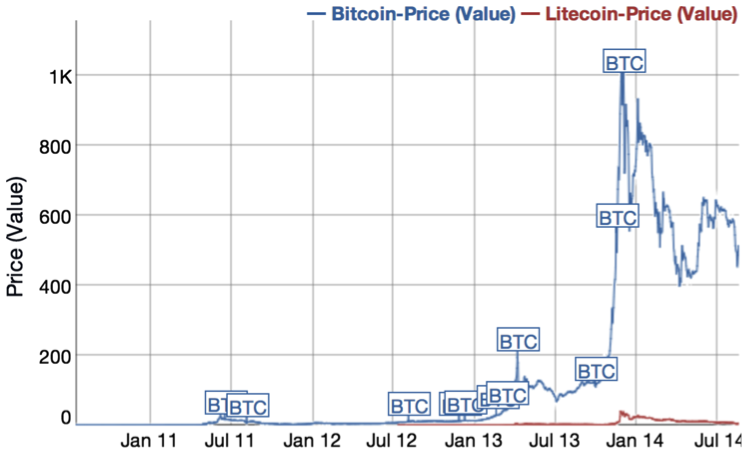}
\caption{Bitcoin and Litecoin Value in USD.}
\end{center}
\end{figure}

I decided to run experiments both mining Bitcoins and Litecoins. Bitcoins use the SHA-256 algorithm, while Litecoins use the Scrypt algorithm. Initially, I tried to distinguish between them, as separate classes. Unfortunately, no good results where achieved, due to almost identical fingerprints, so for the purpose of this article, both Bitcoins and Litecoins have been treated as a single class.\\
For the experiments I used minerd, a CPU miner, needed because of the lack of GPU most servers have. Mining using the CPU is usually slower, but if several big virtual machines perform it, acceptable speeds could be reached.

\subsection{Failures}
To simulate hardware failure I decided to disconnect the network of one of the physical nodes while an instance of HiBench was running in a virtual machine. Several nodes, a master and one or more slaves, which perform the task while the master coordinates them, form a hadoop cluster.
The network failure caused the virtual hadoop node to become isolated from the hadoop cluster, not being able to contact the master, thus terminating the current job as a failure.\\
While running this experiment almost all activity stopped in the physical cluster, including in its virtual machines. It makes sense for this to happen because OpenStack relies on network connectivity to synchronize and manage its services as well as to provide Internet to the virtual resources.

\section{Results}
\subsection{Expected Patterns}
Each one of the experiments should have different characteristics, which make it classifiable. Before trying different classifiers, this section discusses what might make each case unique and if it would be possible to classify them manually, showing the data obtained after the experiments.

\subsection{General workloads}
As previously discussed, two types of general workloads have been run: a Hadoop benchmark and a highly CPU-intensive job. It seems a fair assumption that the hadoop experiment will have values scattered along the entire spectrum, in all the metrics. As I said, HiBench is a collection of jobs that benchmark different aspects of a hadoop cluster, so we would expect, for example, both low and high network usage, as well as CPU and disk. If we look at a scatter plot representing how network and CPU relate to each other as shown in the figure below, we can see that there is no clear pattern.

\begin{figure}[!htb]
\begin{center}
\includegraphics[scale=0.6]{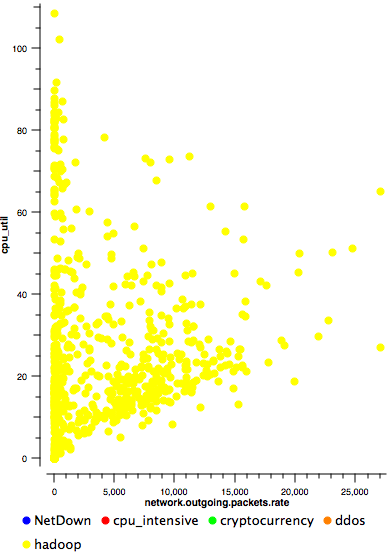}
\caption{Hadoop scatter plot.}
\end{center}
\end{figure}

On the other hand, we have the CPU-intensive job. This job simulates mathematical calculation, which could be taking place in the cloud as means to improve performance and speed. This kind of job would probably have very low, if any, network traffic as well as a relatively low disk usage, in most cases.  However, CPU should be peaking most of the time. Looking at the scatter plot below we can see that the samples cluster on the highest point in CPU usage while staying very low in both network and disk.

\begin{figure}[!htb]
\begin{center}
\includegraphics[scale=0.4]{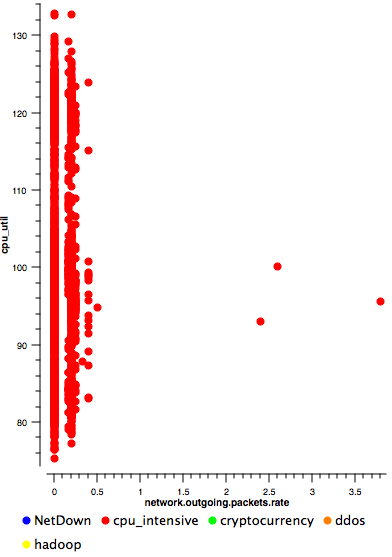}
\includegraphics[scale=0.4]{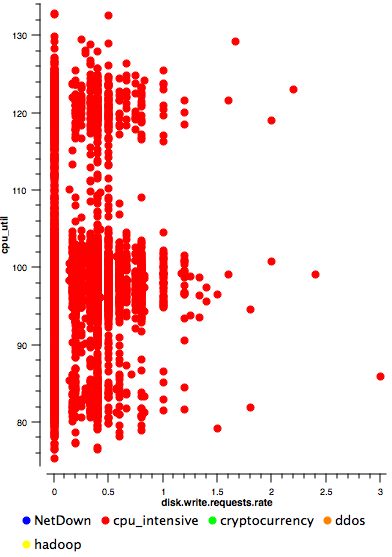}
\caption{CPU intensive scatter plot.}
\end{center}
\end{figure}

\subsection{Fraudulent activity}

\subsubsection{Internal DDoS}
The ping flood attack works by sending a big number of ICMP packets to a victim, using most of its resources, as a consequence. We would expect the data collected to have high values for network and some CPU load, while having relatively low values for disk.\\
As it can be seen in the figure below, a very high number of bytes are being sent while we have a low, but constant CPU usage. Note that we lost some of the data, which appears as several dots in the lowest left part of the plot, possibly due to timeouts caused by the attack.

\begin{figure}[!htb]
\begin{center}
\includegraphics[scale=0.4]{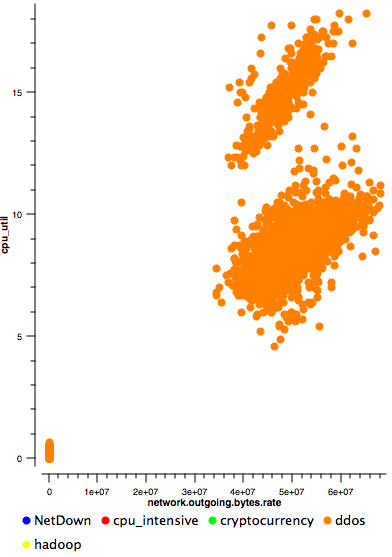}
\caption{DDoS scatter plot.}
\end{center}
\end{figure}

\subsubsection{Cryptocurrency mining}
The process of mining a digital currency usually consists of calculating hashes, what is called proof-of-work, using costly algorithms, which take time and CPU resources. This method is also used for password encryptions in order to render a brute force attack unfeasible. In this experiment I have data of both Bitcoin, using the SHA-256 algorithm, and Litecoin, using the more common Scrypt algorithm.\\
In addition, in order to make the experiment more real, I joined mining pools, so I would also expect a noticeable amount of network traffic present during the mining operation, probably a regular amount of traffic.\\
Looking at the scatter plot below, containing mixed data of Bitcoin and Litecoin generation; we can see that the expected results were correct. The vertical strips visible in the plot show us that there was a repetition of network traffic patterns, which would be correct for cryptocurrency mining.

\begin{figure}[!htb]
\begin{center}
\includegraphics[scale=0.3]{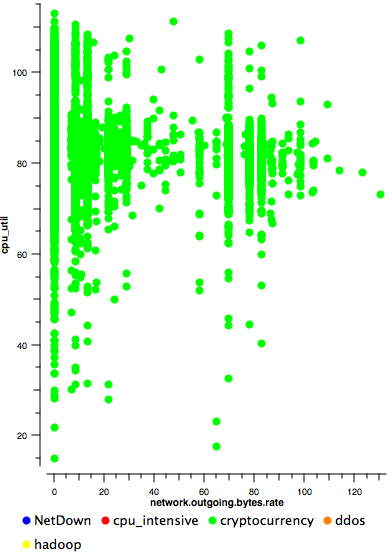}
\caption{Cryptocurrency mining scatter plot.}
\end{center}
\end{figure}

\subsection{Failure}
Finally, for the network down experiment, we would expect to see minimal values of all the metrics, probably making it almost invisible in the plots where several experiments appear together.\\
Indeed, the following scatter plot shows that less than 3% of the CPU is used and, of course, network traffic drops to zero.

\begin{figure}[!htb]
\begin{center}
\includegraphics[scale=0.3]{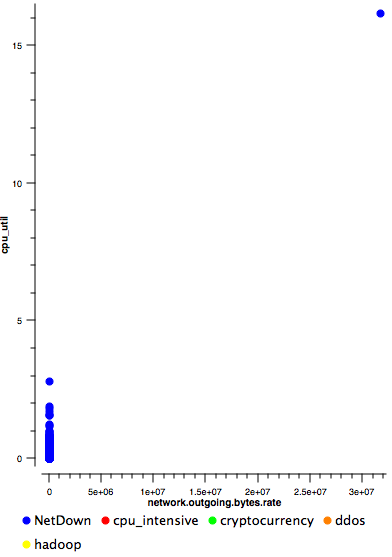}
\caption{Network failure scatter plot.}
\end{center}
\end{figure}

\subsection{Plots}
Having discussed the data we get from each experiment, we also need to discuss whether they are visually distinguishable when represented together. The easier they can be distinguished visually, the easier it will be to classify them. Orange \cite{orange} offers an option to evaluate different representations and return the plot showing us the most relevant information. After running this algorithm, the best ranking representation is the one with CPU utilization and network outgoing packet rate, so let's look at a scatter plot containing all the data collected.

\begin{figure}[!htb]
\begin{center}
\includegraphics[scale=0.4]{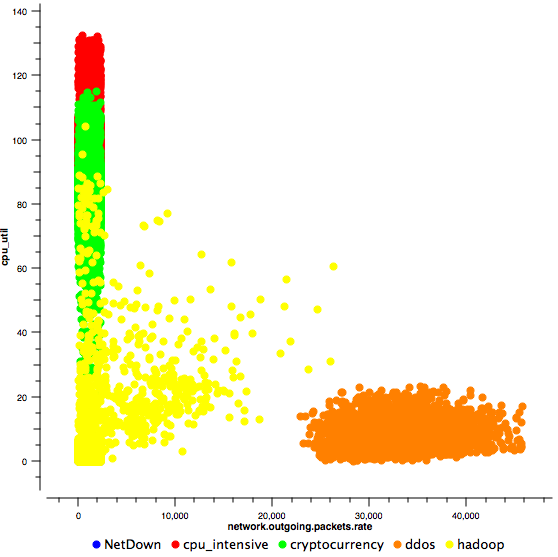}
\caption{Mixed data scatter plot.}
\end{center}
\end{figure}

As it can be seen in the plot above, the data seems to be fairly easy to classify visually. There clearly is some overlapping with the cryptocurrency and CPU-intensive samples and the network failure data does not appear to be visible, overlapping with the hadoop samples. As previously expected, the hadoop data is scattered over the mining and the CPU intensive data. However, we will see if this affects the results of the classification algorithms. Finally, the DDoS samples are completely separated from the rest, which should make it very easy to detect.

\subsection{Algorithm selection}
There are many different classifiers, which are used for many different applications. It is common to try several of them to see which one offers a better result with the kind of data for each problem.\\
In order to rapidly try different algorithms, I used Orange to build a topology, which took the initial data and evaluated several learners, generating reports in order to choose the best one. The topology used was the following:

\begin{figure}[htb]
\begin{center}
\includegraphics[scale=0.5]{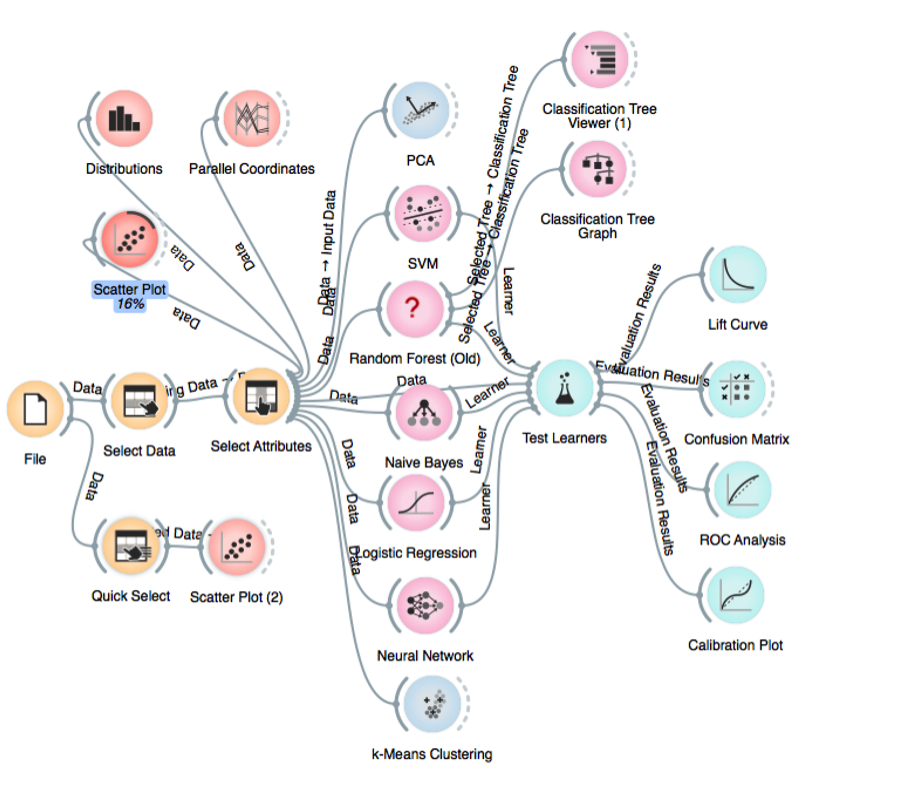}
\caption{Orange topology.}
\end{center}
\end{figure}

As it can be seen in the previous topology, the initial data is loaded and basic transformations are applied to tell Orange which field is the class and which are features. Next, all the classification algorithms are applied to the data. In order to validate the results, cross-validation is performed.\\
Cross-validation is common technique in machine learning to ensure that the results are not coincidence and to ensure they can be replicated. In order to do this, the data we have is divided into folds, being 10 the most used number. Once we have 10 equally sized folds, we use 9 to train the classification model, and the remaining one to evaluate its efficiency. Once the process has finalized, we rotate the folds and repeat it again 10 times, so each fold is used once for evaluation and 9 times for training. As a consequence at the end of the cross-validation process, we can be sure that the results are not coincidence and that they are replicable using different data.\\
Once all the classifiers have finished, the results are sent to the learner tester, which creates statistics and comparisons on performance and accuracy for each classifier. These results can then be plotted or converted to a confusion matrix.\\
The data used contains 36004 instances and, after evaluating it, we will see that the best algorithm was random forest.

\subsection{Analysis}
The following plot represents the average proportion of truths – the percentage of correct classifications – of the algorithms shown in the topology in the previous section.

\begin{figure}[htb]
\begin{center}
\includegraphics[scale=0.4]{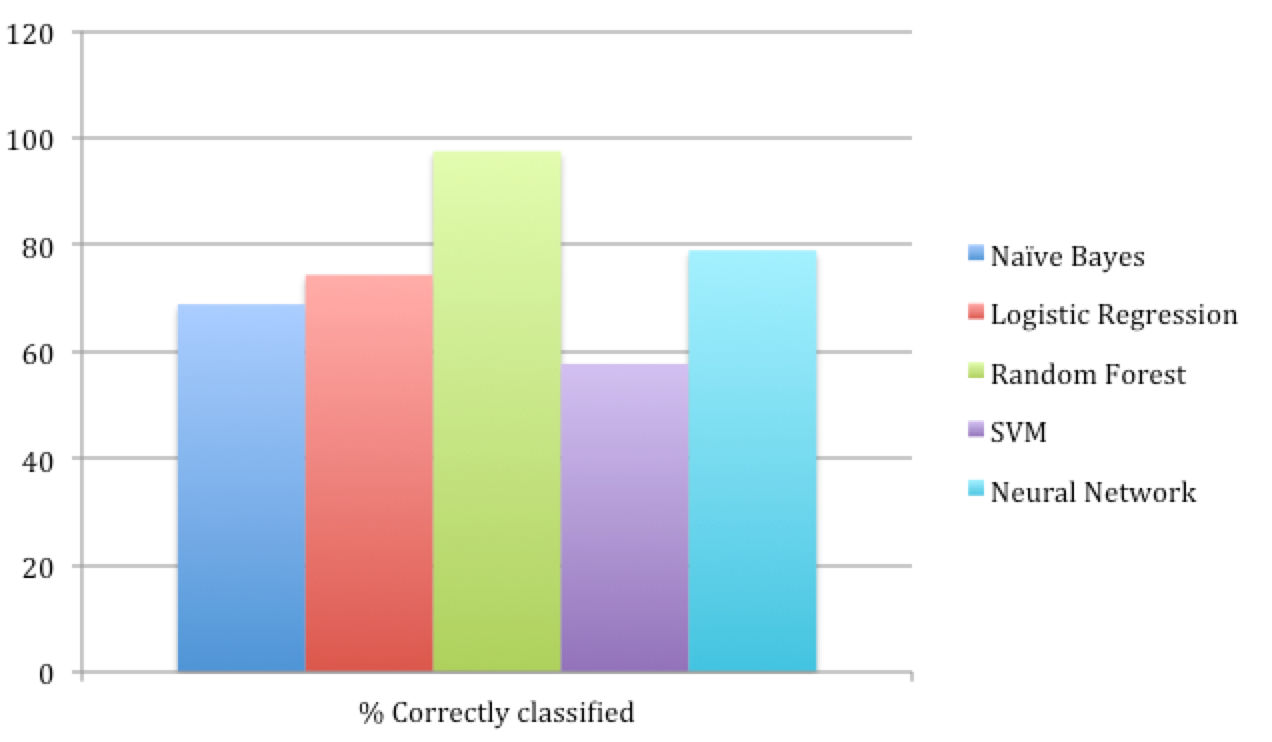}
\caption{Classifier performance comparison.}
\end{center}
\end{figure}

The best one is clearly the random forest classifier with a 97.5\% of samples correctly classified.\\
A random forest is an ensemble of decision trees, which work together. The features of the input data are divided among different trees, depending on how relevant it is for the final choice. The accuracy of the classification can change depending on the number of trees of the forest.\\
Now that the best algorithm had been found, I created a python program that takes the same data as input and applies the random forest classifier from the scikit library. I wrote this, not only to confirm the results provided by Orange but also, to be able to automate trials with more configurations. With it I was able to quickly get performance results for a varying number of trees and with different amounts of folds for cross-validation. These confirmed that the results were correct.\\
Next, I will discuss how the chosen tree looks like and what features seem to be more relevant for a correct classification. This will provide us with a unique insight on the patterns required for identifying these workloads as well as a good understanding of what is happening while a data sample is classified.\\

The chosen tree for the classification is depicted at the figure at the end of the article. The image shows only the first four levels of the tree but it is already enough to see what the most relevant metric is in this case: network.incoming.bytes.rate, which is the root of the tree. According to the condition of whether this value is higher or lower than 50.800 bytes/second, we can already separate most of the samples representing a ddoss or a hadoop job, which represent a higher usage of the network, from the rest.\\
If we look at the left branch, the tree then checks the CPU utilization. In our samples, a hadoop workload shows a lower CPU than a DDoS attack, being able to separate most cases of hadoop samples and more than half of the DDoS ones. Following again the left branch, if the CPU is higher than a threshold, then it might be a cryptocurrency mining or CPU intensive sample. On the other hand, if the CPU is lower than the threshold, we can almost be certain that we are classifying a DDoS sample.\\
Going one step back and looking at the right branch, the tree checks the network.incoming.packets.rate. A DDoS attack will always have a high value for this meter, while a hadoop workload will have a lower one. This condition allows us to classify these two packets with very high accuracy.\\
Going back to the root node and looking at the right branch, we can see that the classification is not as straight forward, and 4 levels are not enough to see what is going on in the tree.\\
These are some important sub-trees that help make decisions to separate classes:

\begin{figure}[!htb]
\begin{center}
\includegraphics[scale=0.7]{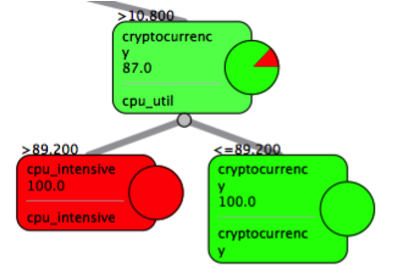}
\caption{Subtree 1.}
\end{center}
\end{figure}

Here we can see some of the most interesting nodes, where cryptocurrency mining is separated from a CPU-intensive activity according to the CPU utilization. It is interesting because classifying between two very CPU intensive workloads using this condition will probably be the cause of some of the incorrect results. This condition is repeated in different places along the tree. In addition, if different workloads are added in the future, the borderline between CPU-intensive allowed activities and cryptocurrency mining will become more blurry in terms of CPU utilization. This justifies the use of a meta-classifier to perform decisions based on several of the last collected samples.

\begin{figure}[!htb]
\begin{center}
\includegraphics[scale=0.6]{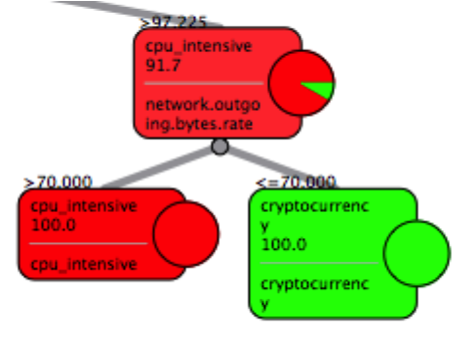}
\caption{Subtree 2.}
\end{center}
\end{figure}

The next example shows how we can distinguish the CPU-intensive workload from the cryptocurrency mining one by looking at the network usage. Indeed, if we do mathematical calculations, the network usage will be much lower than Bitcoin mining, where there is a constant network flow with the mining pool. In this case, if we have more than 70.000 outgoing bytes/second, it decides we are mining a cryptocurrency. Otherwise, it decides we are performing an allowed CPU-intensive activity. Unlike the previous example, this condition is not bound to introduce many errors in the classification, as both jobs have a distinctive network usage fingerprint.

\begin{figure}[!htb]
\begin{center}
\includegraphics[scale=0.7]{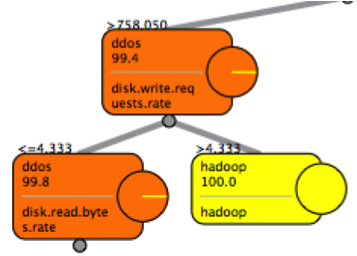}
\caption{Subtree 3.}
\end{center}
\end{figure}

Finally, I'll show an example were the tree classifies between a DDoS and a Hadoop job, represented in the image below, in orange and yellow.\\
When this point in the tree is reached, it means we have very high network utilization and a relatively low CPU utilization. What the tree does in this case is to look at the disk data, which is the metric that makes a difference in this case.\\
As it can be seen, if we have more than a certain number of disk write requests per second (4.333), then we are trying to classify a Hadoop sample. On the other hand, if we have high network usage, relatively low CPU and a low number of disk write requests, we are most likely dealing with a DDoS sample.

\begin{figure}[!htb]
\begin{center}
\includegraphics[scale=0.78]{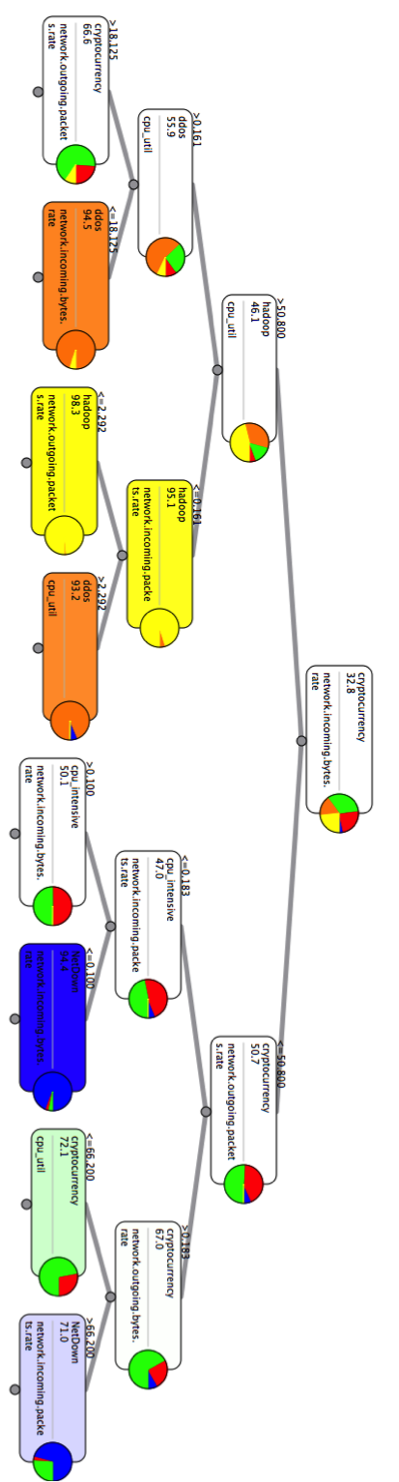}
\caption{Decision tree.}
\end{center}
\end{figure}

\subsection{Meta-classifier}
We now have a classifier that works approximately 97.5\% of the times, however, as the number of workloads increase, this number is bound to decrease. In real life, we cannot expect to detect a Bitcoin mining operation 5 seconds after it starts. Furthermore, it makes sense that such an activity keeps running for a long time, so we want to be able to detect an ongoing fraudulent activity. If we assume that a fair detection interval is one hour, then we can use the individual classifications to perform an hourly statistic and, depending on the overall results, we can decide on a positive or a negative decision.\\
The meta-classifier will look at the results obtained during the last hour. If the percentage of false positives is under a specified threshold, then a positive decision will be made.\\
The previously mentioned python program contains a proof of concept of such a meta-classifier. Given all the classification results calculated in the last hour, plus their correct class, it tries different thresholds to determine which is the minimum error threshold we need with the current model to achieve a 100\% classification success. Surprisingly, with such big classification accuracy, the threshold can be quite low, in fact, as low as 4\% of error would ensure us 100\% accuracy with the current data.\\
In real environments this threshold could be higher or lower depending on how cautious we want to be. If we want to investigate all possible cases, we would set a high error threshold, to make sure that even if we have a 30\% - 40\% of positives in an hour, an alarm gets triggered.\\
On the other hand, if we only want to be noticed when there is a case with a very high probability, we would set the threshold lower, to 5\% - 10\%, for example.\\
The algorithm I used to test the results with different thresholds simulates an hour by taking only the samples that would have been generated in this period of time and assumes that there will be only one virtual machine for each case, to keep it simple. Given this constraint and the correct classification values, it checks if the results from the random forest have an error lower than the threshold, which has been passed as a parameter. This method prints the cases that have not been properly classified or, in case all of them are successfully classified, it prints a success message at the end.

\section{Conclusion}
This article has shown a method to detect fraudulent activities running in our OpenStack cluster by means of a random forest. The data used to detect such activities is formatted of a slice of the last 5 seconds of resource usage for each virtual resource; let it be virtual machines or volumes, among others. As with any method, it offers certain advantages but it also has some shortcomings, which will be discussed in this section.

\subsection{Advantages}
Using data aggregates to classify the activity provides a layer of privacy other methods don't. Of course, detecting a user's activities is not privacy-friendly itself, but with this framework, you do not have detailed information about what they are doing, but are able to detect certain high level patterns, which identify suspicious activity. For example, if a company is using our cloud for confidential activities, we would only be able to know that they are running mathematical calculations, or that they are using distributed systems, which is done by many companies these days. We would not be able, however, to tell what they are calculating or what information is being sent among nodes in distributed systems.\\
Another advantage is the simplicity of collecting this kind of data. Imagine we are a web hosting company and we run on top of OpenStack. Then we will boot virtual machines to be used as web server by our customers. Let's also assume that the cloud provider does not offer an intrusion detection system. Using the method proposed here, even without being administrators of the cloud, not having direct access to the physical resources data, we can still obtain the required data from our virtual machines and send it to our own systems for offline, or even real time, analysis.

\subsection{Shortcomings}
As with any method, there are shortcomings to using it and it has some limitations.\\
The most important shortcoming is the amount of lost information in the data aggregates. While it does not affect much with the tested activities, as the number of different classes to classify increases, so will the number of similar workloads, decreasing the effectiveness of the classification. This is why this method can be used of a first step of a fraudulent activity detection pipeline. It can be used to detect suspicious activity and then trigger an alarm that starts a more in-depth IDS that will determine if it was truly a fraudulent workload or not. By doing this, we reduce the amount of data an IDS has to process, while being able to reuse the data that is collected in order to bill the customer.

\subsection{Future work}
Thanks to the use of OpenStack, this article has been accepted as a conference talk in the next OpenStack Summit, which takes places in Paris, this November. The talk, which will be a more OpenStack-focused, is called “Using Ceilometer data to detect fraudulent activity in our OpenStack cluster” \cite{summit} and several people have already emailed the author expressing their interest in watching it.\\
In terms of development, the next step is to create a real time pipeline that runs in a production ready OpenStack cluster, classifying the activity, using a meta-classifier, and sending hourly reports. In order to do this, an easy way would be to modify Ceilometer so it performs the classification for each sample and saves the result in the database as a new metric. Then, once an hour, the meta-classifier can retrieve the last samples and perform the final decision.\\
This method, modifying Ceilometer, makes it easy to create a real time pipeline for OpenStack. However, this would not be compatible with other clouds, such as AWS or Microsoft Azure. It would probably be better to create everything as a separate program that accepts data input via a restful API. This would allow the classification itself to take place anywhere, even in the cloud itself. In addition, it would allow cloud users, not administrators, to monitor their own resources, having complete control of this data.



%%\bibliographystyle{elsarticle-num} 
%%\nocite{*}
%%\bibliography{bib.bib}


\end{document}